\documentstyle[aps,multicol,epsf,prl]{revtex} %from YK%
\newcommand{\r}  { {\bf r}}
\newcommand{\e}  { {\rm e}}
\newcommand{\dr} { {\rm d}\r\ }

\begin{document}
\title{Steric Effects in Electrolytes: A Modified Poisson--Boltzmann
Equation}
\author{Itamar Borukhov and  David Andelman}
\address{School of Physics and Astronomy, 
Raymond and Beverly Sackler Faculty of Exact Sciences,
Tel Aviv University, Tel Aviv 69978, Israel}
%\and
\author{Henri Orland}
\address{Service de Physique Th\'eorique, CE-Saclay,
     91191 Gif-sur-Yvette, Cedex, France}
\date{\today}
\maketitle

\begin{abstract}
  The adsorption of large ions from solution to a charged surface is
investigated theoretically.  A generalized Poisson--Boltzmann equation,
which takes into account the finite size of the ions is presented.  We
obtain analytical expressions for the electrostatic potential and ion
concentrations at the surface, leading to a modified Grahame equation. At
high surface charge densities the ionic concentration saturates to its
maximum value.  Our results are in agreement with recent experiments. 
 \end{abstract}
\draft

\begin{multicols}{2} %from YK

%\section{Introduction}

The interaction between charged objects (interfaces, colloidal particles,
membranes, etc) in solution is strongly affected by the presence of an
electrolyte (salt)  and is of great importance in biological systems and
industrial applications \cite{israelachvili,hunter}.  The main effect is
screening of the Coulomb interaction characterized by the so-called
Debye--H\"uckel screening length \cite{debye}, which depends on the ionic
strength of the solution. The Deryaguin-Landau-Verwey-Overbeek theory,
based on the competition between screened Coulomb and attractive van der
Waals interactions, has been very successful in explaining the
stabilization of charged colloidal particles \cite{DLVO}.

One of the most widely used analytical method to describe electrolyte
solutions is the Poisson--Boltzmann (PB) approach \cite{GC}. For low
electrostatic potentials (less than 25 mV), the PB equation can be
linearized and yields the Debye-H\"uckel theory \cite{debye}. The PB is a
continuum mean-field like approach assuming point-like ions in
thermodynamic equilibrium and neglecting statistical correlations. This
theory has been successful in predicting ionic profiles close to planar
and curved surfaces and the resulting forces.  However, it is known to
strongly overestimate ionic concentrations close to charged surfaces.  In
particular, this shortcoming of the PB theory is pronounced for highly
charged surfaces and multivalent ions.

%motivation

Since the PB equation does not take into account the finite size of the
adsorbing ions, the ionic concentration close to the surface can easily
exceed the maximal allowed coverage by orders of magnitude. Several
attempts have been proposed to include the {\it steric} repulsion in order
to improve upon the PB approach \cite{henderson,mirkin}. One of the first
attempts to incorporate steric effects is the {\it Stern layer}
modification \cite{henderson,stern} of the PB approach. Steric effects are
introduced by excluding the ions from the first molecular layer close to
the surface. However, it seems difficult to improve on this method in a
systematic way.  More recent modifications
\cite{henderson,mirkin,HNC,MSA,parsegian} rely either on Monte Carlo
computer simulations or on numerical solutions of integral equations (the
``hypernetted chain'' equation \cite{HNC}). These approaches involve
elaborate numerical calculations and lack the simplicity of the original
PB approach. 

In this Letter, we propose a simple way to include steric effects in the
original PB approach. This modified PB equation clearly shows how ionic
saturation takes place close to a charged surface. The equation is derived
for 1:$z$ asymmetric and $z$:$z$ symmetric electrolytes. At low ionic
concentration, the original PB equation is recovered.  Simple analytical
relations between the surface charge density and the counterion
concentration at the surface are obtained, in agreement with recent
experiments \cite{rondelez}. 

%\section{The Mean--Field Equation}

Consider an asymmetric electrolyte consisting of negative multivalent ions
of charge $-z e$, and positive monovalent ions of charge $e$, where $e$ is
the electron charge.  The bulk concentration of the negative and positive
ions is $c_b$ and $z c_b$, respectively, as implied by charge neutrality. 

Within mean--field approximation, the total free energy, $F=U-TS$, can be
written \cite{epl} in terms of the local electrostatic potential
$\psi(\r)$ and the ion concentrations $c^\pm(\r)$.  The electrostatic
energy contribution $U$ is: 
 \begin{eqnarray}
  \label{Energy}
  U &=& \int\dr \biggl[-{\varepsilon\over 8\pi}|\nabla\psi|^2
    + ec^+\psi - zec^-\psi \nonumber\\
  & &   -\mu_+c^+  -\mu_-c^- 
               \biggr]
\end{eqnarray}
  The first term is the self energy of the electric field, where
$\varepsilon$ is the dielectric constant of the solution.  The next two
terms are the electrostatic energies of the ions, and the last two terms
couple the system to a bulk reservoir, where $\mu_\pm$ are the chemical
potentials of the ions. 

\noindent
The entropic contribution $-TS$ is
\begin{eqnarray}
  \label{Entropy}
  -TS &=& {k_BT \over a^3}  \int\dr
         \biggl[ {c^+a^3}\ln\bigr(c^+a^3\bigr) +
                 {c^-a^3}\ln\bigl(c^-a^3\bigr) \nonumber\\
      & &+ ~~\bigl(1-c^+a^3-c^-a^3\bigr) 
           \ln\bigl(1-c^+a^3-c^-a^3\bigr) \biggr]
\end{eqnarray}
 where $k_BT$ is the thermal energy.  For simplicity, we assume that both
types of ions have the same size $a$.  The first two terms are the
entropies of the positive and negative ions, whereas the last term is the
entropy of the solvent molecules. Indeed, this last term is responsible
for the novel steric corrections to the PB equation. In a more rigorous
way, these corrections are obtained by considering a lattice-gas version
of the Coulomb gas in which each lattice site is occupied at most by one
ion \cite{tobepublished}. 

  The variation of the free energy $F=U-TS$ with respect to $\psi$ and
$c^\pm$ yields our modified PB equation for the 1:$z$ electrolyte: 
 \begin{eqnarray}
   \nabla^2\psi &=& - {4\pi\over \varepsilon} 
    \bigl[ec^+(\r) - zec^-(\r) \bigr] \nonumber \\ 
    &=& {4\pi z e c_b \over\varepsilon}
   {\e^{z\beta e\psi} - \e^{-\beta e\psi} \over
    1 - \phi_0 + \phi_0 (\e^{z\beta e\psi} + z\e^{-\beta e\psi})/(z+1)}
  \nonumber \\   
  \label{PB1toz} 
\end{eqnarray}
   where $\phi_0 = (z+1) a^3 c_b$ is the total bulk volume fraction of the
positive and negative ions. 

For a symmetric
$z$:$z$ electrolyte, one gets
\begin{eqnarray}
   \nabla^2\psi &=& {8\pi ze c_b\over\varepsilon} 
   {\sinh(z\beta e\psi) \over
    1 - \phi_0 + \phi_0 \cosh(z\beta e\psi)}
   \label{PBztoz}
\end{eqnarray}
 where $\phi_0 = 2 a^3 c_b$.  In the limit of small ionic concentrations,
$\phi_0\to 0$, Eqs. \ref{PB1toz},\ref{PBztoz} reduce to the standard PB
equations.  Moreover, for any ionic concentration and at low electrostatic
potentials, $|\beta e\psi|\ll 1$, both equations reduce to the linearized
PB equation (Debye-H\"uckel limit)  $\nabla^2\psi = \kappa^2\psi $ where
$\kappa^{-1}$ is the Debye--H\"uckel screening length.  For the asymmetric
case $\kappa^2 = 4\pi l_b z(z+1) c_b$, where $l_b=e^2/\varepsilon k_BT$ is
the Bjerrum length equal to $7$\AA\ for aqueous solutions at room
temperature. 

Our approach deviates significantly from the original PB equation for
large electrostatic potentials $|\beta e \psi|\gg 1$.  In particular, the
ionic concentration is unbound in the standard PB approach, whereas here
it is always bound by $1/a^3$ (``close packing'') as can be seen from Eqs.
\ref{PB1toz},\ref{PBztoz}.  This effect is important close to strongly
charged surfaces immersed in an electrolyte solution. 

Note that for high positive potentials, $\beta e \psi \gg 1$, the
contribution of the positive ions is negligible and the negative ion
concentration follows a distribution reminiscent of the Fermi-Dirac
distribution \cite{kornyshev},
 \begin{eqnarray}
c^-(\r) \to {1 \over a^3}~~\frac{1}{1+
(z+1){{1-\phi_0}\over\phi_0}\e^{-z\beta e\psi}}
  \label{FermiDirac}  
\end{eqnarray}
 where the excluded volume interaction plays the role of the Pauli
principle. 

To demonstrate the usefulness of our method, we study the case of a single
planar surface with charge density $\sigma>0$ in contact with an
electrolyte solution. Ionic concentration profiles are obtained from the
numerical solution of Eq.~\ref{PB1toz} as a function of $x$, the distance
to the positively charged surface.  Since the positive ion concentration
is small near the surface, we show in Fig.~1a only the negative ion
profiles, as well as the corresponding original PB profile. The main
effect is the saturation of the ionic concentration in the vicinity of the
charged surface.  This should be contrasted with the original PB scheme
which leads to extremely high and unphysical values of $c^-_s\equiv
c^-(0)$, especially for multivalent ions.  In the saturated region, the
ionic concentration tends to $1/a^3$, leading to more pronounced
deviations from PB for large ions. 

In the saturated layer the right--hand side of Eq.~\ref{PB1toz} becomes a
constant, and the electrostatic potential is quadratic
 \begin{eqnarray}
\label{psi-strong}
\psi(x) \simeq \psi_s - {4\pi\sigma\over\varepsilon}x
     + {2\pi ze\over\varepsilon a^3} x^2 
\end{eqnarray}
 where $\psi_s$ is the surface potential and the boundary condition
$\psi'|_s = -4\pi\sigma/\varepsilon$ is satisfied.  As can be seen in
Fig.~1b, the parabolic profile of $\psi(x)$ is a good approximation close
to the surface. The width of the saturated layer $l^*$ is not strictly
equal to $a$. It can be easily estimated from Eq.~\ref{psi-strong} to be
$l^*\simeq a^3\sigma/z e$ in agreement with Fig.~1a. 

 The surface potential $\psi_s$ can be calculated in a closed form from
the first integral of Eq.~\ref{PB1toz}, assuming that the concentration of
the positive ions is negligible at the surface
\begin{eqnarray}
  \psi_s &=& {k_B T\over ze}\biggl\{
      \ln\Bigl[ \e^\zeta -\bigl(1-\phi_0 \bigr) \Bigr] 
    - \ln\bigl(c_ba^3\bigr) \biggr\} 
  \nonumber \\
  &\approx&{k_B T\over ze}
          \biggl\{ \zeta - \ln\bigl(c_ba^3\bigr) \biggr\}
  \label{psi-s}
\end{eqnarray}
   where 
\begin{eqnarray}
  \label{zeta}
  \zeta \equiv  {2\pi a^3\sigma^2\over \varepsilon k_B T}
\end{eqnarray}

  Similarly, the concentration of negative ions at the surface 
can be calculated leading to a modified Grahame Equation 
\cite{israelachvili}
\begin{eqnarray}
  \label{cs}
  c^-_s = {1\over a^3}\Bigl[1- (1-\phi_0)\e^{-\zeta}\Bigr]
\end{eqnarray}
  This contribution is depicted in Fig.~2a, where $c^-_s$ is plotted as a
function of the surface charge density, $\sigma/e$, for two different ion
sizes, $a$.  The PB case is shown as well for comparison.  At low surface
charge $\zeta\ll 1$, the ion concentration reduces to the PB results
 \begin{eqnarray}
  \label{cs-pb}
  c^-_s = {2\pi\sigma^2 \over \varepsilon k_BT} + (1+z)c_b
 \end{eqnarray}
but for high surface charge $\zeta \gg 1$, the deviations
from the PB case are substantial. Furthermore, as can be seen from
the above equation, the ionic concentration near the surface
depends only weakly on the bulk electrolyte concentration, $c_b$.

In Fig.~2b the ratio between the ion charge density of the first layer
$\sigma_1 \simeq ze c^-_s a$ and $\sigma$ is plotted as function of the
specific surface area per unit charge, as is often measured in
experiments.  For the PB approach, this ratio diverges at high surface
charge densities (small specific area) because $c^-_s \sim \sigma^2 $. 
However, in our case, the steric effect changes the situation altogether
since it prevents the ions from approaching and over-compensating the
surface charges.

The theoretical results presented here are relevant to recent experiments
\cite{rondelez} where large multivalent ions are adsorbed onto a charged
Langmuir monolayer.  Large polyanions such as H$_3$PW$_{12}$O$_{40}$
(phosphotungstic acid) dissolved in an aqueous subphase are attracted to a
cationic Langmuir monolayer such as a fatty amine surfactant
(C$_{20}$H$_{41}$-NH$_2$), spread at the water/air interface.  The
adsorbed ion density (per unit area) in the solution, $\sigma_1$, is
measured by X-ray reflectivity.  It is then related to the surface charge
density $\sigma$, which is controlled by Langmuir trough. The experiments
show very clearly the presence of the steric effects for these large ions
(of estimated size of 10\AA). As the surface charge density increases,
$\sigma_1/\sigma$ decreases in accord with our findings (Fig.~2b) and in
contrast to the original PB approach.

In conclusion, we have derived a modified PB equation including steric
effects. As a result the ionic concentration cannot exceed a saturation
value of $1/a^3$. This effect is in particular important for large ions
adsorbing on charged surfaces.  We have obtained analytical expressions
for the potential and ion concentrations at the surface.  Our results
differ qualitatively and quantitatively from the standard PB equation and
agree with recent experiments on large polyions in solution. 

It would be interesting to further explore the connection between our
analytical approach and the Stern layer approach, as well as look at the
implications on the forces between two charged planar surfaces. 
  
%\acknowledgments{
 We wish to thank N. Cuvillier and F. Rondelez for communicating their
experimental results to us prior to publication and for useful
discussions.  We also benefited from discussions and correspondence with
H. Diamant, M. E. Fisher, A. A. Kornyshev, P. Sens and M. Urbakh.  Partial
support from the US-Israel Binational Foundation (BSF) under grant No.
94-00291 is gratefully acknowledged.  One of us (HO) would like to thank
the Sackler Institute of Solid State Physics (Tel Aviv University)  for a
travel grant.

\end{multicols}

\section*{Figure Caption}

\noindent {\bf Fig.~1}:
 (a) Concentration profiles of negative multivalent ions $c^-(x)$ near a
positively charged surface as obtained from the numerical solution of
Eq.~\ref{PB1toz} for two different ion sizes $a=7.5$\AA\ and $a=10$\AA.
Note that the saturated layer width is $l^*\simeq2$\AA\ and $5$\AA,
respectively.  The solid line represents the concentration profile of the
standard PB equation. 
 (b) Calculated electrostatic potential profiles near the surface plotted
together with the parabolic approximation (Eqs.~\ref{psi-strong},
\ref{psi-s}).  The dotted, dashed and solid lines are as in (a).  The bulk
concentration is $c_b=0.1$M for a 1:$z$ electrolyte with $z$=4. The
surface charge density $\sigma$ is taken as one electron charge per
50\AA$^2$. The aqueous solution with $\varepsilon=80$ is at room
temperature. 

\noindent {\bf Fig.~2}:
 (a) Surface concentration of ions as a function of the surface charge
density for different ions size, $a$.  The PB concentration is also
plotted for comparison. 
 (b) Ratio of the first layer charge density and the surface charge
density, $\sigma_1/\sigma=ze c^-_s a/\sigma$, as a function of the
specific surface area per unit charge, $e/\sigma$, for different ion
sizes. The PB result is plotted with the first layer width taken as
$5$\AA.  The 1:$z$ electrolyte bulk concentration is $c_b=1m$M and the
valency $z=4$.  
%\newpage

\begin{figure}[tbh]
\epsfxsize=0.5\linewidth
\centerline{\hbox{ \epsffile{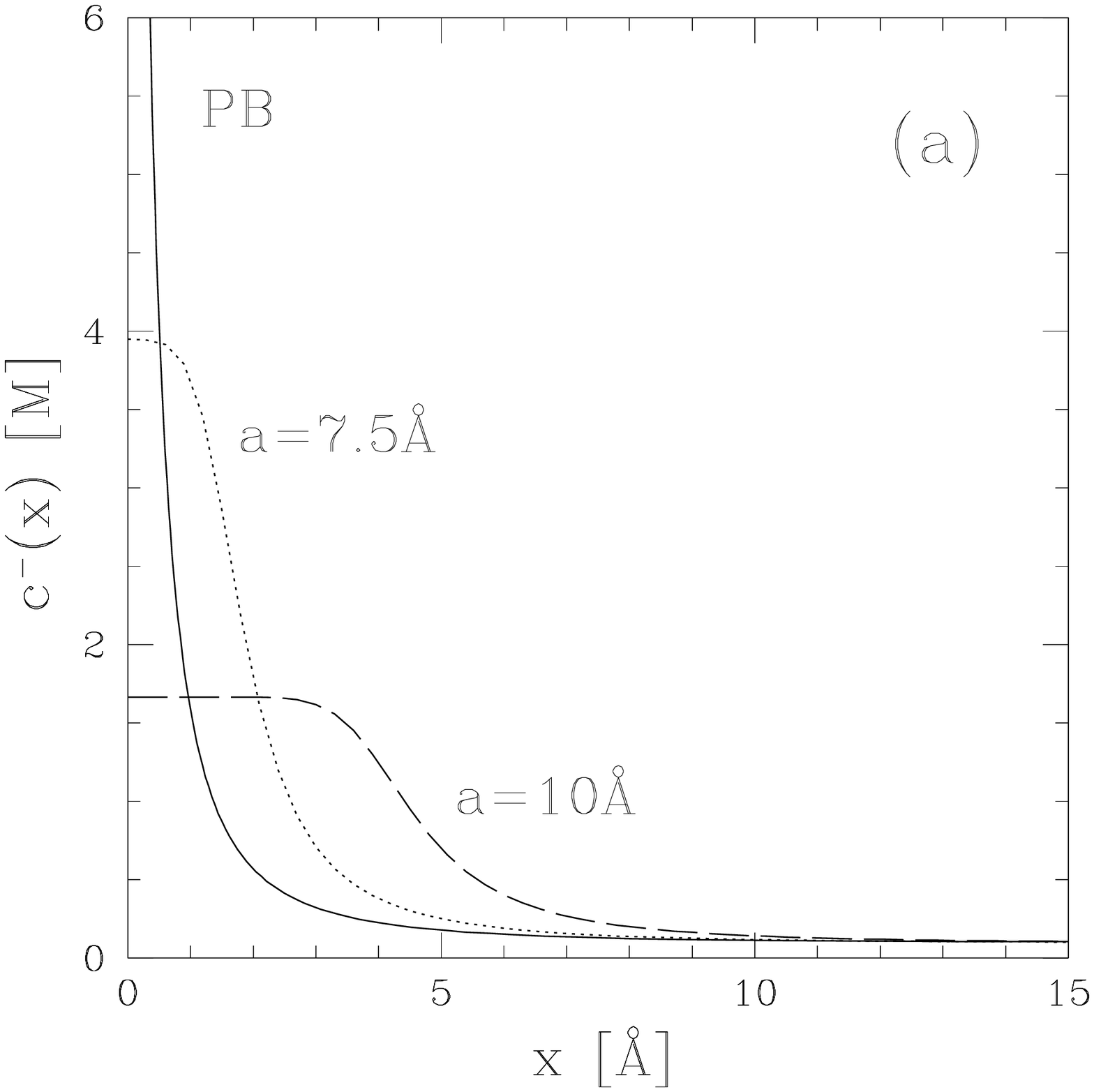} } 
\epsfxsize=0.5\linewidth
            \hbox{ \epsffile{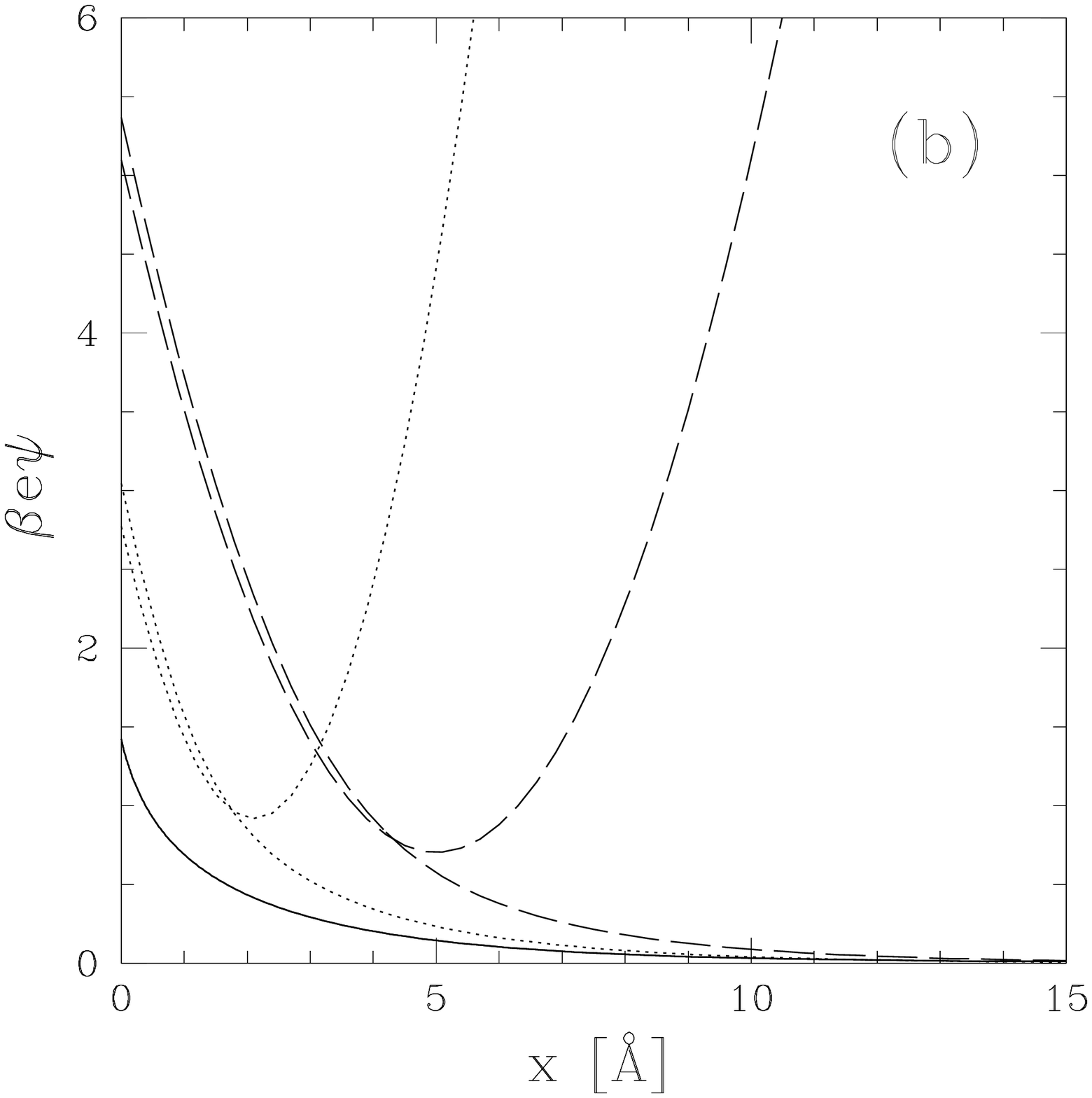} } }
\end{figure}
\centerline{\large Fig.~1}

\begin{figure}[tbh]
\epsfxsize=0.5\linewidth
\centerline{\hbox{ \epsffile{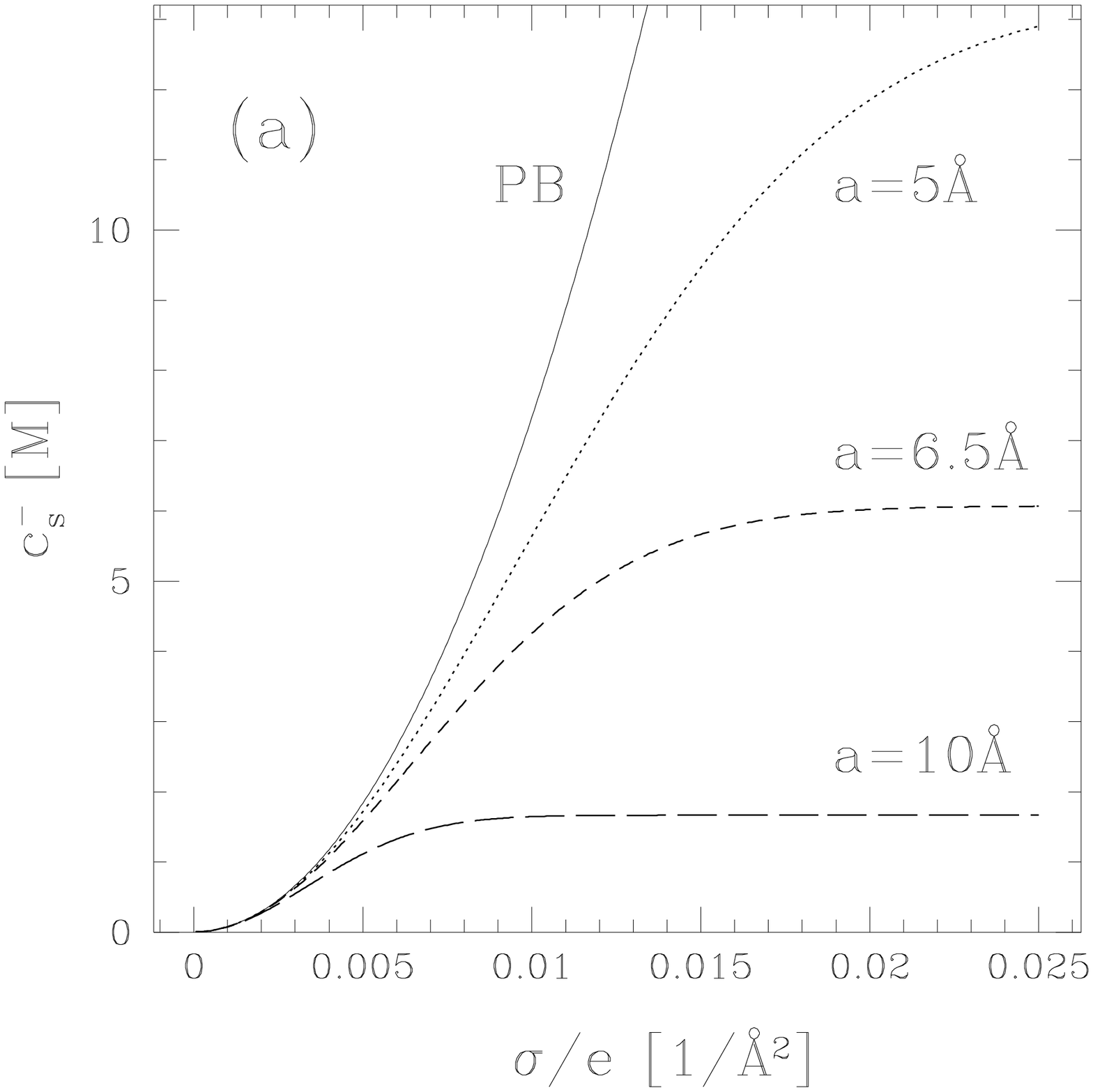} }  
\epsfxsize=0.5\linewidth
            \hbox{ \epsffile{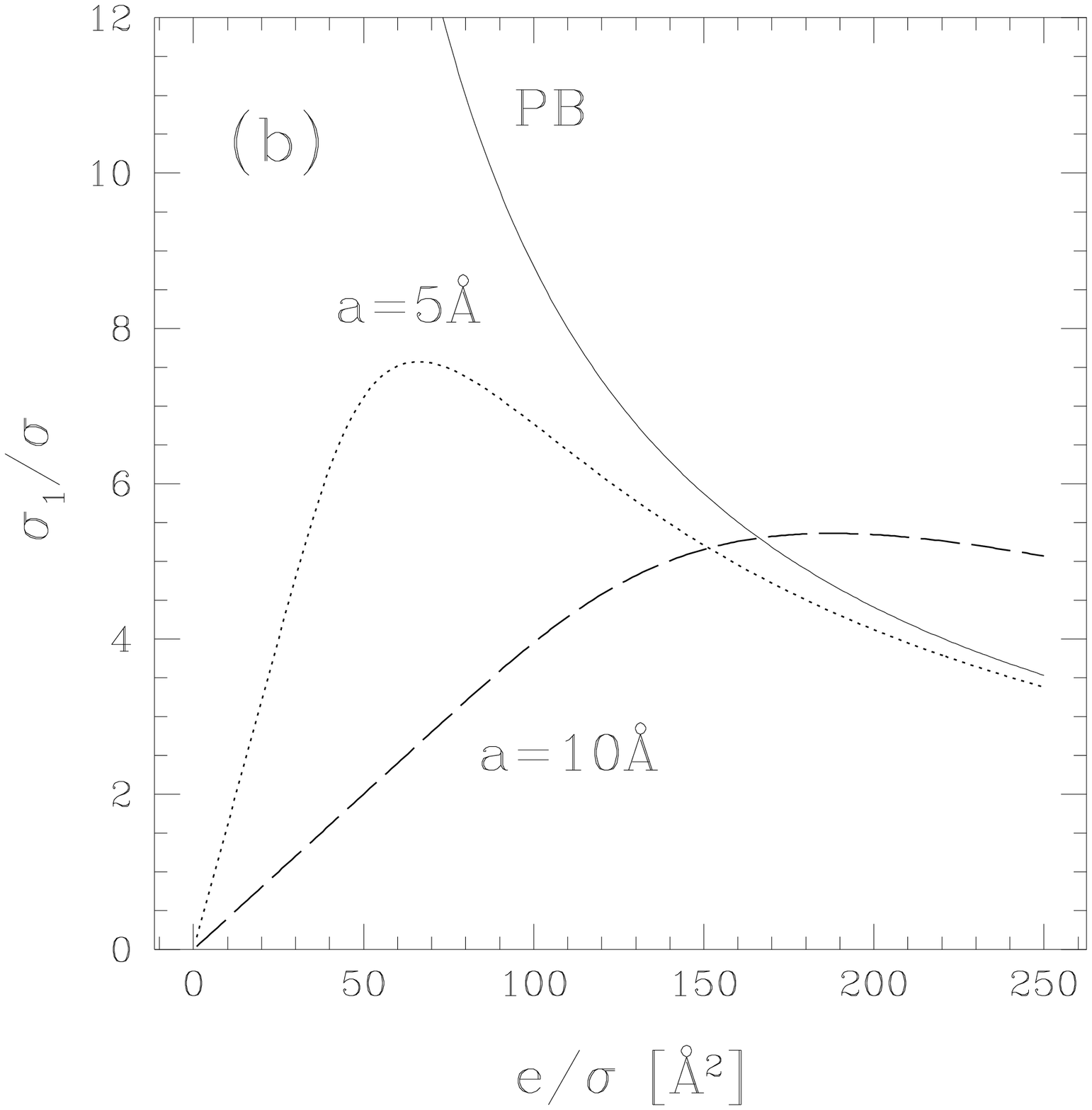} } }
\end{figure}
\centerline{\large Fig.~2}

\end{document}